\documentclass[twocolumn,showpacs,preprintnumbers,amsmath,amssymb]{revtex4-1}
\usepackage{graphicx}% Include figure files
\usepackage{bm}% bold math
\usepackage{float}

\begin{document}

\title{Half-metallic Antiferromagnet Sheets in Sr$_{4}$M$_{2}$O$_{6}$CrFeAs$_{2}$
(M=Sc, Cr) and Their Bulk Form}

\author{Shu-Jun Hu and Xiao Hu}
\affiliation{ World Premier International Center for Materials Nanoarchitectonics (MANA)\\
National Institute for Materials Science, Tsukuba 305-0044, Japan}

\date{\today}

\begin{abstract}
We reveal by first-principles calculations that in iron pnictides
Sr$_{4}$M$_{2}$O$_{6}$CrFeAs$_{2}$ (M=Sc, Cr) the two-dimensional CrFeAs$_{2}$
layers exhibit a robust band structure of half-metallic
antiferromagnet (HMAFM). Due to the thick blocking layer,
the interlayer coupling is vanishingly small and thus the conductive
channels in individual layers may take the alternative spin
direction randomly. We show that, since the spin
magnetizations of Fe and Cr are different in a ferromagnetic state,
applying a strong magnetic field and ramping it down gradually can
align the spin direction of conductive channel of all HMAFM layers,
which restores the bulk HMAFM.
\end{abstract}

%Pnictides (non-cuprate superconductors), 74.70.Xa
%spin polarized transport devices, 85.75.-d
%Electronic structure ab initio calculations, 31.15.A-

\maketitle

%\section{Introduction}
%spin -> half-metal
Spin is one of the fundamental properties of electron. Exploitation
of the spin-dependent transportation, dubbed as spintronics, which
emerged from the discovery of giant magnetoresistance effect
\cite{gmrgrunberg,gmrfert}, has been a topic of particular interest
in the last decades. Generation of the spin-polarized current is the
primary requirement of spintronics devices. Half metals
\cite{halfmetal}, a class of materials possess states in only one
spin channel at the Fermi level ($E_{\rm F}$), have received
considerable attention since they yield fully
spin-polarized currents.

%half-metal -> HMAFM
Owing to the asymmetric electron populations in the two spin
channels, half metals are often ferromagnetic (FM) or ferrimagnetic.
Half-metallic antiferromagnet (HMAFM) materials proposed first by
van Leuken and de Groot\cite{deGroot} should be superior
in many applications, where stray fields are harmful, since they achieve
fully spin-polarized currents without showing net magnetization. To date,
many HMAFM candidates have been
engineered by the first-principles calculations based on the double
perovskite compounds \cite{exmaple-dp-1,exmaple-dp-2,exmaple-dp-3}
and Heusler alloys \cite{exmaple-heusler}.
Besides, HMAFM have been predicted by doping in
oxide semiconductor \cite{dms}, transition-metal (TM)
oxides \cite{tm-oxide}and cuprate\cite{cuprate}.

In a previous study\cite{BaCrFeAs2}, the present authors proposed to
achieve HMAFM based on the recently renaissant iron pnictides \cite{Hosono}.
We substituted half of the Fe atoms by Cr in BaFe$_{2}$As$_{2}$.
The ground state was characterized by the following properties: the Fe
and Cr atoms form intervening three-dimensional (3D) lattice, the spin
magnetizations of Fe, Cr and As compensate completely, and close to
$E_{\rm F}$ the states are contributed exclusively from the spin
channel of Cr:3d majority electrons. The material BaFe$_{2}$As$_{2}$
is therefore predicted as a HMAFM.

\begin{figure}
\includegraphics[width=8.5cm]{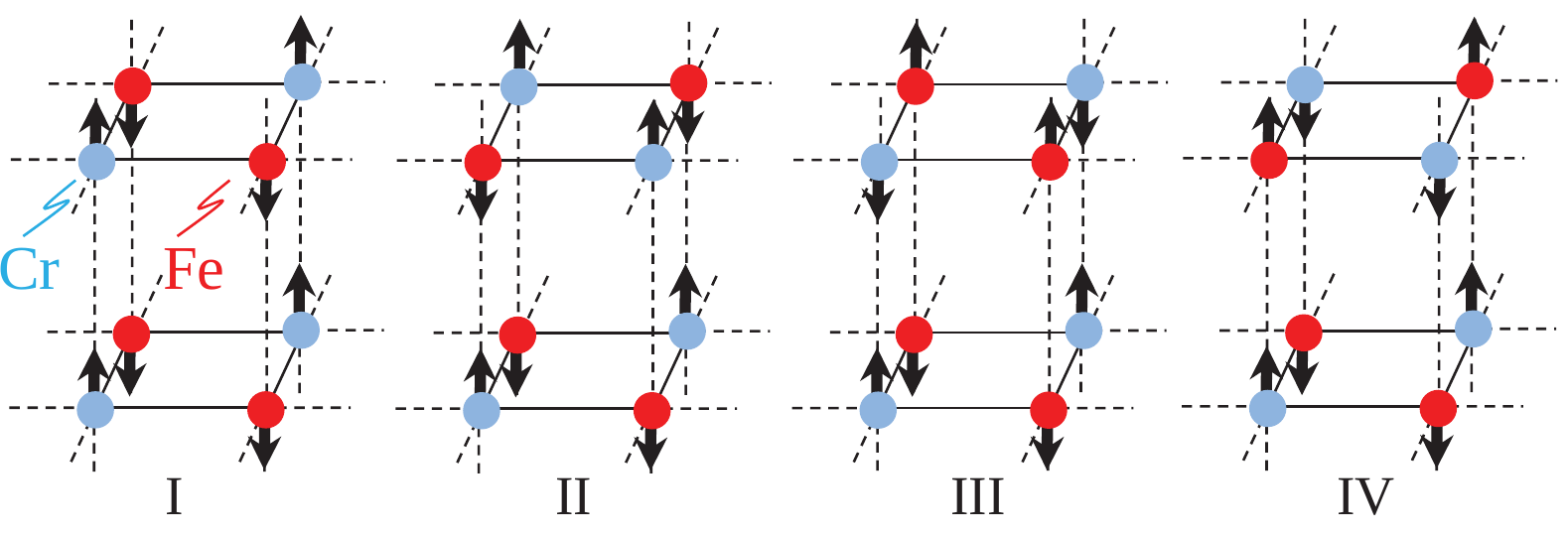}
\caption{\label{lowest} (Color online) Lattice and
spin orders of four degenerated configurations of TM layers in the ground state
of Sr$_{4}$Sc$_{2}$O$_{6}$CrFeAs$_{2}$.
The As ions and blocking layers in the unit cell are omitted for a clear view.
Configuration I and II, in which all Cr (and Fe) ions take the same spin direction,
are predicted to be HMAFM, while III and IV are not.}
\end{figure}

In the present work, we focus on the recently discovered iron
pnicitides Sr$_{4}$M$_{2}$O$_{6}$Fe$_{2}$As$_{2}$ (M=Sc, Cr) \cite{TBL-Kishio}. These new
iron pnictides (see also \cite{TBL-Wen-1,TBL-Wen-2})
are unique in the structure where the FeAs layers are separated
by layers of perovskite structure as thick as 9.8 $\AA$ , and thus
is much more two-dimensional (2D) than other iron pnictides. We substitute
half of the Fe atoms by the conjugate element Cr as revealed in
Ref.\cite{BaCrFeAs2}. From first-principles calculations it becomes
clear that in the ground state the individual TM pnictide
(CrFeAs$_{2}$) layers take the checkerboard order of Cr and Fe ions
associated with the antiferromagnetic (AFM) spin order, and exhibit HMAFM
electronic structure. Due to the vanishing interlayer coupling,
both lattice and spin disorders easily appear in
the \emph{c} direction, which suppresses the bulk HMAFM property. Interestingly,
it is found that the spin magnetizations of Cr
ions are much larger than Fe ions in the FM state. Therefore,
a post-process of applying a strong
magnetic field and ramping down gradually can restore the spin order
along the \emph{c} direction and thus achieve the 3D HMAFM.
The present approach to achieve the bulk property based on the
robust 2D one prepared in advance may shed light for future
exploration of HMAFM.

\begin{figure}
\includegraphics[width=8.5cm]{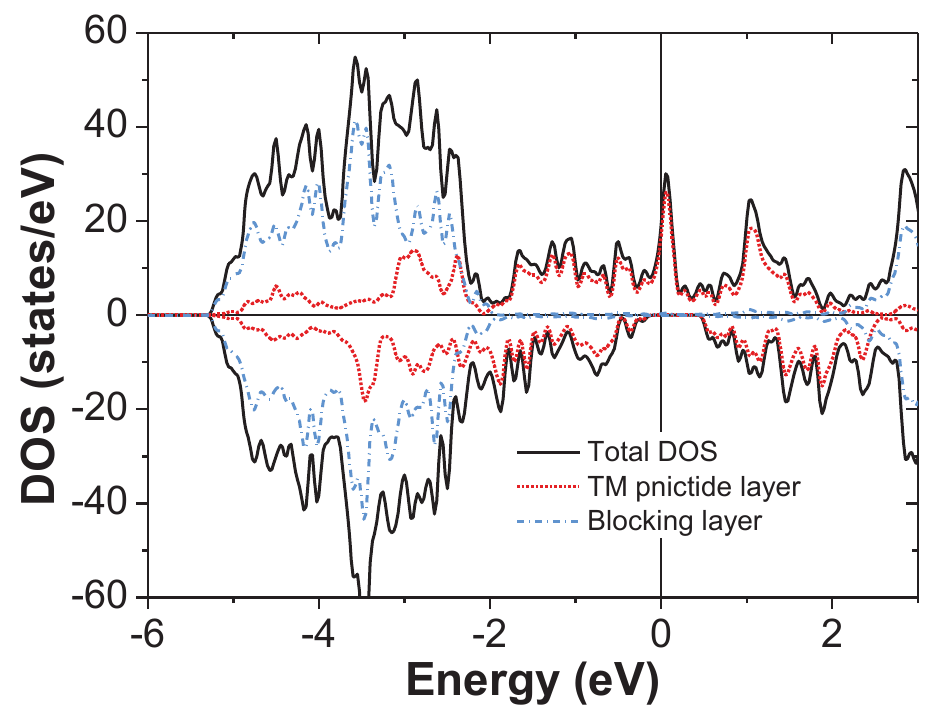}
\caption{(Color online) Total density of states (DOS)
and those projected on the TM pnictide layer and blocking
layer of Sr$_{4}$Sc$_{2}$O$_{6}$CrFeAs$_{2}$,
respectively, for the atomic and spin orders I and II in Fig.~\ref{lowest}.}
\label{dosfigure}
\end{figure}

%\section{Method and Results}

%\subsection{Calculation Details}
The calculations are performed using the projector augmented-wave
(PAW) method \cite{PAW-method} as implemented in the VASP simulation
package \cite{vasp-cite-1,vasp-cite-2}, with the generalized
gradient approximation in PBE type \cite{PBE} for the
exchange-correlation functionals. The crystal structures of the
parent materials revealed
experimentally \cite{TBL-Kishio} are implemented in the present
calculations for the doped system.  Structure optimization
is then performed using 8$\times$8$\times$2 Monckhorst-Pack
\cite{monkhorst-pack} k-point mesh
until the stress of the supercell and the force of the
ions are less than 1 kbar and 10$^{-3}$ eV/$\AA$.

%\subsection{Sr$_{4}$Sc$_{2}$O$_{6}$CrFeAs$_{2}$}

We begin with the identification of ground state of
Sr$_{4}$Sc$_{2}$O$_{6}$CrFeAs$_{2}$ by calculating the total
energies of possible configurations within the
$\sqrt{2}\times\sqrt{2}\times2$ supercell. The checkerboard
configuration of Cr and Fe ions associated with AFM
spin magnetizations is of energy lower than other
2D configurations at least by 0.23 eV.
The robust in-plane checkerboard order is consistent with
the case of BaCrFeAs$_{2}$ \cite{BaCrFeAs2}, but in sharp
contrast with the study for the '1111' system \cite{nakao-1111}.

%table:
%conf., Cr, Fe, As, Total
\begin{table}
\caption{\label{magnetization} Spin magnetizations of Cr, Fe and As
ions and total spin magnetization (a) in states I and II shown
in Fig.~\ref{lowest} and (b) in the FM state (units: $\mu _{\textrm B}$).
The total spin magnetization is obtained by integrating
charge densities of both spin channels.}
\begin{ruledtabular} %double line
\begin{tabular}{ccccc}
Conf. & Cr & Fe & As & M$_{tot}$ \\
  \hline
(a)   & 2.83 & -2.68 & -0.08 & 0.00  \\
(b)   & 2.45 &  0.56 & -0.08 &11.33  \\
\end{tabular}
\end{ruledtabular}
\end{table}

On the other hand, due to the diminished coupling across the thick
blocking layer, the four configurations shown in Fig.~\ref{lowest} with
different orders along the c axis possess almost the same energy
($\Delta E\le 3$ meV).

Figure~\ref{dosfigure} shows the total density of states (DOS) and those
projected to the block layer and CrFeAs$_{2}$ layer, for the configurations I
and II in Fig.~\ref{lowest}, where the Cr (Fe) 3d majority electrons are spin-up (spin-down).
The band near $E_{\rm F}$ (from -2 to 1 eV) is dominated by states
from CrFeAs$_{2}$ layers, and all
the occupied states from the blocking layer are deep in the valence
band. This result reveals the role of blocking layer as the charge
reservoir. At $E_{\rm F}$ the DOS is exclusively contributed by
the spin-up states, which consist of the Cr:3d majority and Fe:3d minority
electrons, and a part of As:4p states. As illustrated in Table~\ref{magnetization},
the total magnetization is zero due to the complete
compensation of the spin magnetizations of Cr, Fe and As ions. Therefore,
the material is a HMAFM in the states I and II shown in Fig.~\ref{lowest}.

\begin{figure}
\includegraphics[width=8.5cm]{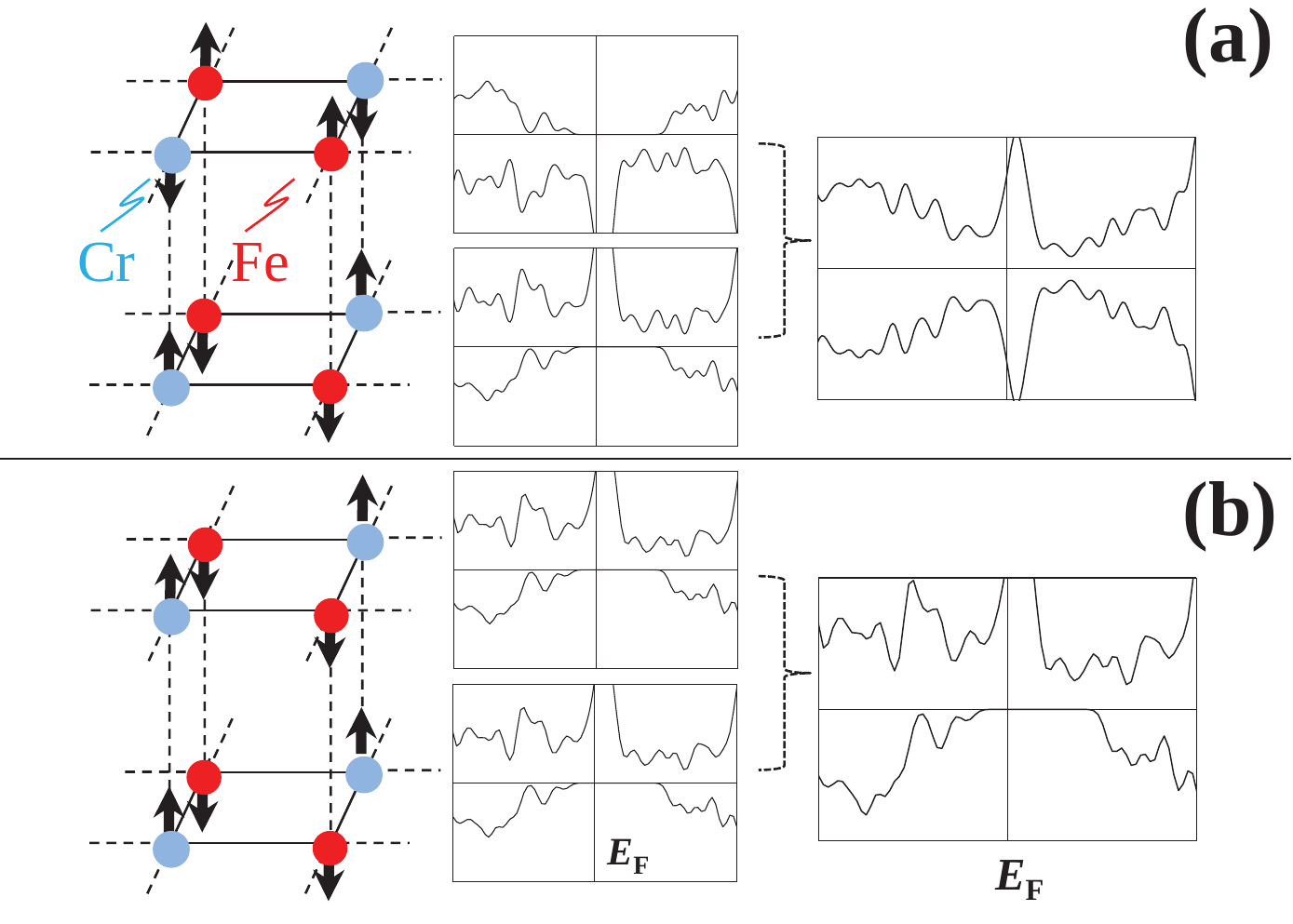} \caption{(Color online) From (a)
the state III in Fig.~\ref{lowest} (2D HMAFMs) to (b) a
bulk HMAFM. Left column: lattice and spin orders; center column:
DOSs of individual TM layers; right column: total DOS. DOSs are
shown in energy range from -1 to 1 eV relative to $E_{\rm F}$.}
\label{schematicfigure}
\end{figure}

In BaCrFeAs$_{2}$ \cite{BaCrFeAs2}, the strong interlayer coupling via the direct As bonding
delocalizes the As:4p states, resulting a narrow band gap of 0.3 eV.
In Sr$_{4}$Sc$_{2}$O$_{6}$CrFeAs$_{2}$, the coupling is diminished due to the
thick blocking layer.
Therefore, the As:4p states are well confined in the CrFeAs$_{2}$ layer.
The energy gap of Sr$_{4}$Sc$_{2}$O$_{6}$CrFeAs$_{2}$ is calculated to be 0.7 eV,
which is much larger than that of BaCrFeAs$_{2}$.

For the configuration III in Fig.~\ref{lowest}, the total DOS is
symmetric in the two spin channels, as shown in the right column of
Fig.~\ref{schematicfigure}(a). However,
projecting the total DOS on the blocking layers and the two individual
TM pnictide layers, we find that, first, the blocking layers still do not contribute
to states close to $E_{\rm F}$, and secondly, the two TM pnictide
layers are HMAFMs with the metallic channels in opposite spin directions,
as shown in the center column of Fig.~\ref{schematicfigure}(a). The same results are obtained
for configuration IV in Fig.~\ref{lowest}.
Therefore, Sr$_{4}$Sc$_{2}$O$_{6}$CrFeAs$_{2}$ shows a robust 2D HMAFM, while the bulk
HMAFM may be suppressed due to the degenerated states of different spin orders
along the \emph{c} axis.

Since the spin order can be tuned by external magnetic field, we investigate
the property of FM spin configuration,
which is achieved when the external magnetic field is strong enough. It is revealed
by first-principles calculations that in the FM state the spin magnetization
of Cr is much larger than that of Fe as displayed in Table~\ref{magnetization},
in a sharp contrast to the AFM state where they are almost the same. The
difference in the spin magentizations, and thus in the Zeeman energies,
can be used to tune the spin orders. In a strong magnetic field overcoming
the energy difference between the FM and AFM states, the spin magnetizations
of both Cr and Fe are simply aligned along the external field. During the
process of ramping the external field down to zero, the spin
magnetization of Fe flips to the direction opposite to the external field,
since the spin magnetization of Fe presumes smaller Zeeman energy and
the AFM state is much more stable than the FM state. Thus, no matter what the
initial c-axis spin order is, the AFM order between Cr and Fe can
be restored over the bulk material as shown in Fig.~\ref{schematicfigure}(b).
The 3D HMAFM property is eventually achieved.

\begin{figure}
\includegraphics[width=8.5cm]{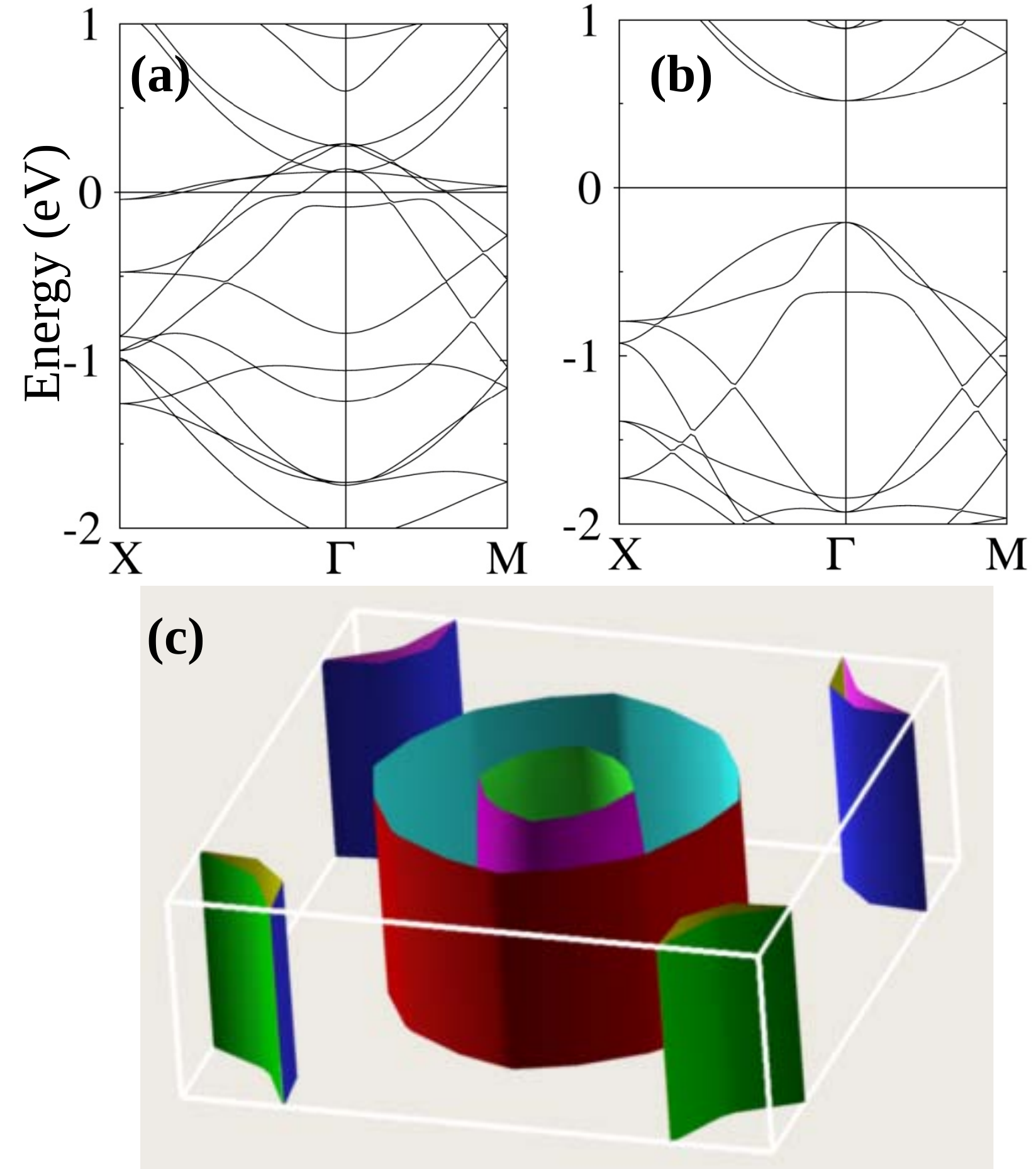}
\caption{(Color online) Band structure of
Sr$_{4}$Sc$_{2}$O$_{6}$CrFeAs$_{2}$. (a) spin-up band;
(b) spin-down band. (c) Fermi surfaces of the spin-up band.}
\label{bandfigure}
\end{figure}

%band structure
We then look at the band structure of 3D HMAFM
Sr$_{4}$Sc$_{2}$O$_{6}$CrFeAs$_{2}$. For the configuration I in
Fig.~\ref{lowest}, we use a $\sqrt{2}\times\sqrt{2}\times1$
supercell to perform the calculation. As displayed in
Fig.~\ref{bandfigure}(a), the partially occupied bands in the
spin-up channel include two hole bands contributed by the
TM:t$_{2g}$ electrons and two almost degenerated
Fe:d$_{3z^{2}-r^{2}}$ bands corresponding to the peak in DOS near
$E_{\rm F}$ in Fig.~\ref{dosfigure}. The Fermi surfaces of the
spin-up channel are shown in Fig.~\ref{bandfigure}(c). The hole
pockets near the $\Gamma$ point are induced by the TM:t$_{2g}$
states and two Fermi surfaces near the X points of the Brillouin
zone are contributed by the Fe:d$_{3z^{2}-r^{2}}$ states. All the
Fermi surfaces show a more significant 2D character as compared with
the iron-pnictides with thinner blocking layers
\cite{fermi-surface}. There is no Fermi surface in the spin down
channel since a gap is opened.

First-principles calculations reveal that the band
structure of configuration II is almost the same as that of
configuration I, both characterized by the same Cr (Fe) spin
direction over the bulk. Therefore, we predict that, as an
aggregate of configurations I and II, the bulk
Sr$_{4}$Sc$_{2}$O$_{6}$CrFeAs$_{2}$ subjected to the
processing of an external field is of the same
band structure as shown in Fig.~\ref{bandfigure}.

%math.exp(-0.41*1.6*10**(-19)/(1.38*10**(-23)*1500))
Next let us address whether Cr atoms in Sr$_{4}$Sc$_{2}$O$_{6}$CrFeAs$_{2}$ would go to the
blocking layers. In the blocking layer, only Sc atoms might be
substituted by Cr, which forms the Cr$_{Sc}$-Sc$_{Cr}$ anti-site
defects. Using a $2\sqrt{2}\times2\sqrt{2}\times1$ supercell we calculate
the formation energy of anti-site defect, which is estimated to be as large
as 2.41 eV. The Cr$_{Sc}$-Sc$_{Cr}$ anti-site
defect is thus effectively prohibited even the synthesizing
temperature reaches 1500 K \cite{TBL-Kishio}. Therefore, the doped Cr
ions prefer exclusively the Fe sites.

\begin{figure}
\includegraphics[width=8.5cm]{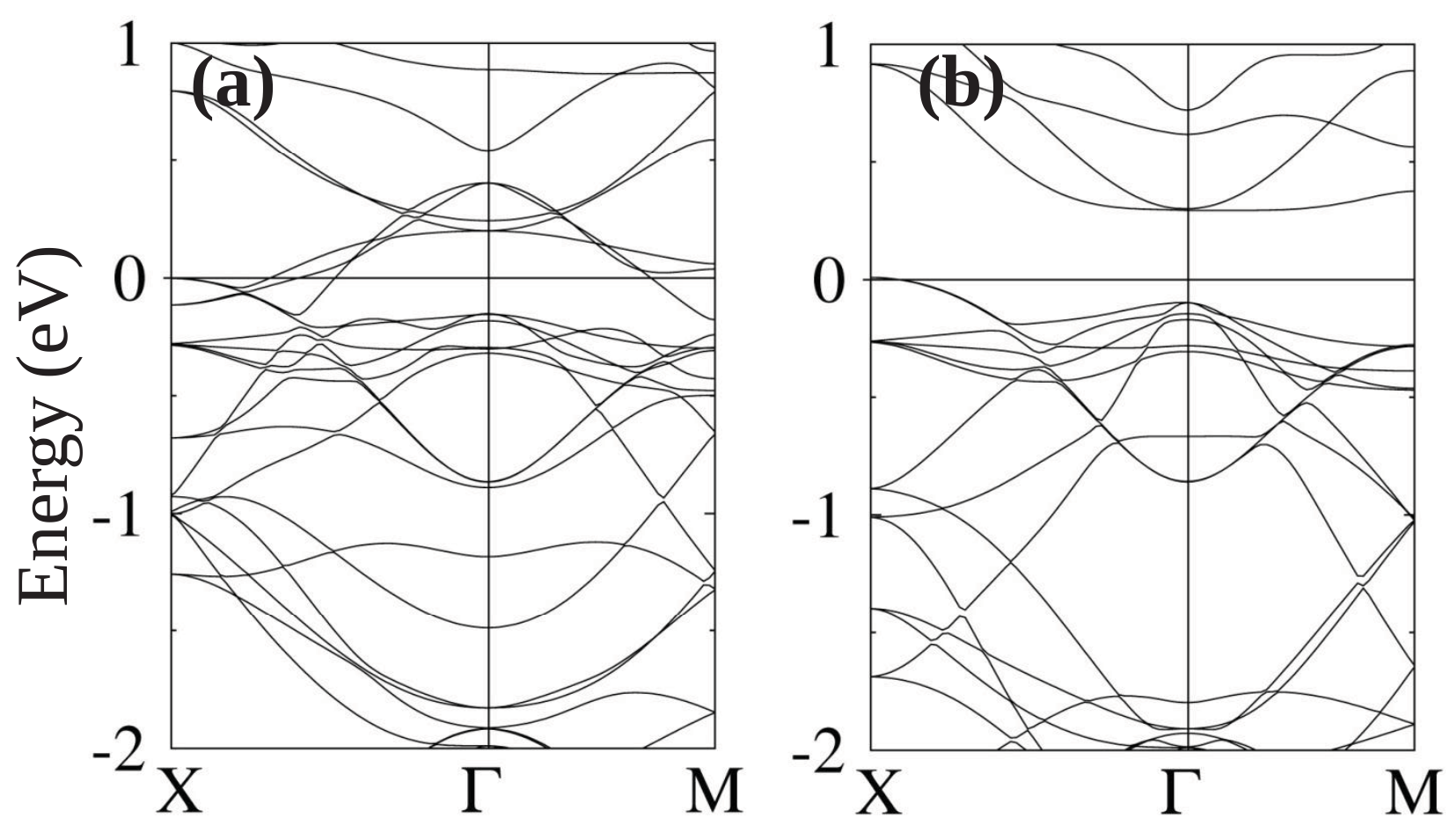}
\caption{ Band structure of
Sr$_{4}$Cr$_{2}$O$_{6}$CrFeAs$_{2}$. (a) spin-up band; (b) spin-down band.}
\label{morecrfigure}
\end{figure}

%\subsection{Sr$_{4}$Cr$_{2}$O$_{6}$CrFeAs$_{2}$}
Finally we turn to the twin parent material
Sr$_{4}$Cr$_{2}$O$_{6}$Fe$_{2}$As$_{2}$, where the Cr atoms
exclusively occupy the B-sites of perovskite structure in the
blocking layer \cite{TBL-Kishio}.
In order to manipulate the band structures, we try the same approach,
namely replacing half of the Fe atoms with Cr atoms in the FeAs
layers. We find that in CrFeAs$_{2}$ sheets the Fe and Cr atoms
prefer the checkerboard distribution associated with an AFM order
among Fe and Cr ions same as the case of
Sr$_{4}$Cr$_{2}$O$_{6}$CrFeAs$_{2}$, and that Cr ions in the
blocking layers form AFM order themselves consistently with the
experimental observation \cite{SrCrOFeAs-spin-order}. As shown in
Fig.~\ref{morecrfigure}, the material
Sr$_{4}$Cr$_{2}$O$_{6}$CrFeAs$_{2}$ has a gap in the spin-down
channel and is metallic in the spin-up channel, and thus is predicted to be another
HMAFM. The gap is 0.2 eV, smaller than the previous case since Cr
atoms in the blocking layers yield additional bands around $E_{\rm
F}$. In the majority band, the orbitals d$_{yz}$, d$_{xz}$ and
d$_{x^{2}-y^{2}}$ of Cr atoms are fully occupied, while the other two orbitals
are well above $E_{\rm F}$ due to the Coulomb repulsion between them and p
electrons of oxygen; all 3d orbitals in the minority band are
empty.

%\section{Discussion and Conclusion}

To summarize, we reveal by first-principles calculations that in
iron pnictides Sr$_{4}$Sc$_{2}$O$_{6}$CrFeAs$_{2}$
and Sr$_{4}$Cr$_{2}$O$_{6}$CrFeAs$_{2}$ the CrFeAs$_{2}$
layers exhibit a robust, two-dimensional property of half-metallic
antiferromagnet. Although the weak interlayer coupling may allow
spin disorders along the c axis, the antiferromagnetic spin order
between Cr and Fe can be restored over the bulk material by applying
a pulse of external magnetic field, which gives rise to
three-dimensional half-metallic antiferromagnet.

\vspace{3mm} \noindent {\it Acknowledgements -- } The calculations were
performed on Numerical Materials Simulator (SGI Altix) of NIMS.  This work was
supported by WPI Initiative on Materials Nanoarchitectonics, MEXT, and
partially by Grants-in-Aid for Scientific Research (No.22540377), Japan.

\end{document}